\documentclass{edm_article}
\begin{document}


\title{Can Population-based Engagement\\ Improve Personalisation?\\ A Novel Dataset and Experiments}

\numberofauthors{1}
\author{
Sahan Bulathwela$^1$, Meghana Verma$^2$, Mar\'ia P\'erez-Ortiz$^1$, Emine Yilmaz$^1$ and John Shawe-Taylor$^1$\\
       \affaddr{$^1$ Centre for Artificial Intelligence, University College London, UK}\\
       \affaddr{$^2$ Indian Institute of Technology Bombay, India}\\
       \email{m.bulathwela@ucl.ac.uk}
}

\maketitle
\begin{abstract}
This work explores how population-based engagement prediction can address cold-start at scale in large learning resource collections. 
The paper introduces i) VLE, a novel dataset that consists of content and video based features extracted from publicly available scientific video lectures coupled with implicit and explicit signals related to learner engagement, ii) two standard tasks related to predicting and ranking context-agnostic engagement in video lectures with preliminary baselines and iii) a set of experiments that validate the usefulness of the proposed dataset. Our experimental results indicate that the newly proposed VLE dataset leads to building context-agnostic engagement prediction models that are significantly performant than ones based on previous datasets, mainly attributing to the increase of training examples. VLE dataset's suitability in building models towards Computer Science/ Artificial Intelligence education focused on e-learning/ MOOC use-cases is also evidenced. Further experiments in combining the built model with a personalising algorithm show promising improvements in addressing the cold-start problem encountered in educational recommenders. This is the largest and most diverse publicly available dataset to our knowledge that deals with learner engagement prediction tasks. The dataset, helper tools, descriptive statistics and example code snippets are available publicly.

\end{abstract}

\keywords{Population-based Engagement, Cold-start, Educational Recommender, Personalised Education, AI in Education}

\section{Introduction}
With the rapid growth of Open Educational Resources (OER) \cite{unesco1} and Massively Open Online Courses (MOOC) \cite{ramesh2014learning,Guo_vid_prod}, large educational resource repositories need scalable tools to understand and estimate the engagement potential of newly added materials \cite{clementsQ}. 
While \emph{contextualised engagement} can be defined as learner's engagement driven by variables related to the context of the learner at a given time in their learning path (e.g., learning needs/goals, knowledge state, background on the topic, etc.), \emph{context-agnostic engagement} aims to capture patterns and features associated with engagement that instead are applicable to an entire learner population rather than individual contexts of specific learners \cite{context_agnostic_engagement}. Put simply, context-agnostic engagement is concerned with the features that generally make a piece of educational material engaging. 


For works such as this, it is important to clarify that 
Information Retrieval (IR) in education has very distinct characteristics than conventional web-searches, as users tend to \emph{gather} information. Learners discovering knowledge in an educational IR system usually carry out learning and familiarising with novel concepts during the search session leading to the task deviating or expanding into unanticipated sub tasks \cite{cortinovis2019supporting}. This behaviour is drastically different from typical search engine users who are usually aware of the exact information they are after. 
This distinction has led to many works that are specific to educational IR \cite{DavisSHH18,PenhaH20}. This difference also makes conventional query log datasets sub-optimal when training e-learning IR models. Prior works have shown this mismatch \cite{Chen18} leading to using education specific datasets  \cite{Syed2017} instead of general-purpose query logs. 


This work attempts to enrich the educational IR research domain by making a large educational video dataset available to the community. \textbf{Video Lecture Engagement (VLE)} is a novel dataset that presents around 12,000 peer-reviewed scientific video lectures constructed from a popular OER repository, VideoLectures.NET and contains a variety of lecture types ranging from scientific talks to expert panels to MOOC-like lectures. The majority of these lectures belong to Computer Science (CS), Artificial Intelligence (AI) and Data Science subject areas, making this dataset a great source for understanding learner engagement with AI/CS related educational videos. The dataset provides an extensive set of textual and video-specific features extracted from the lecture transcripts, together with Wikipedia topics covered in the lecture (via entity linking) and user engagement labels (both explicit and implicit) for each video. 

\begin{figure}[!tbp] 
  \centering
  \includegraphics[width=.7\linewidth]{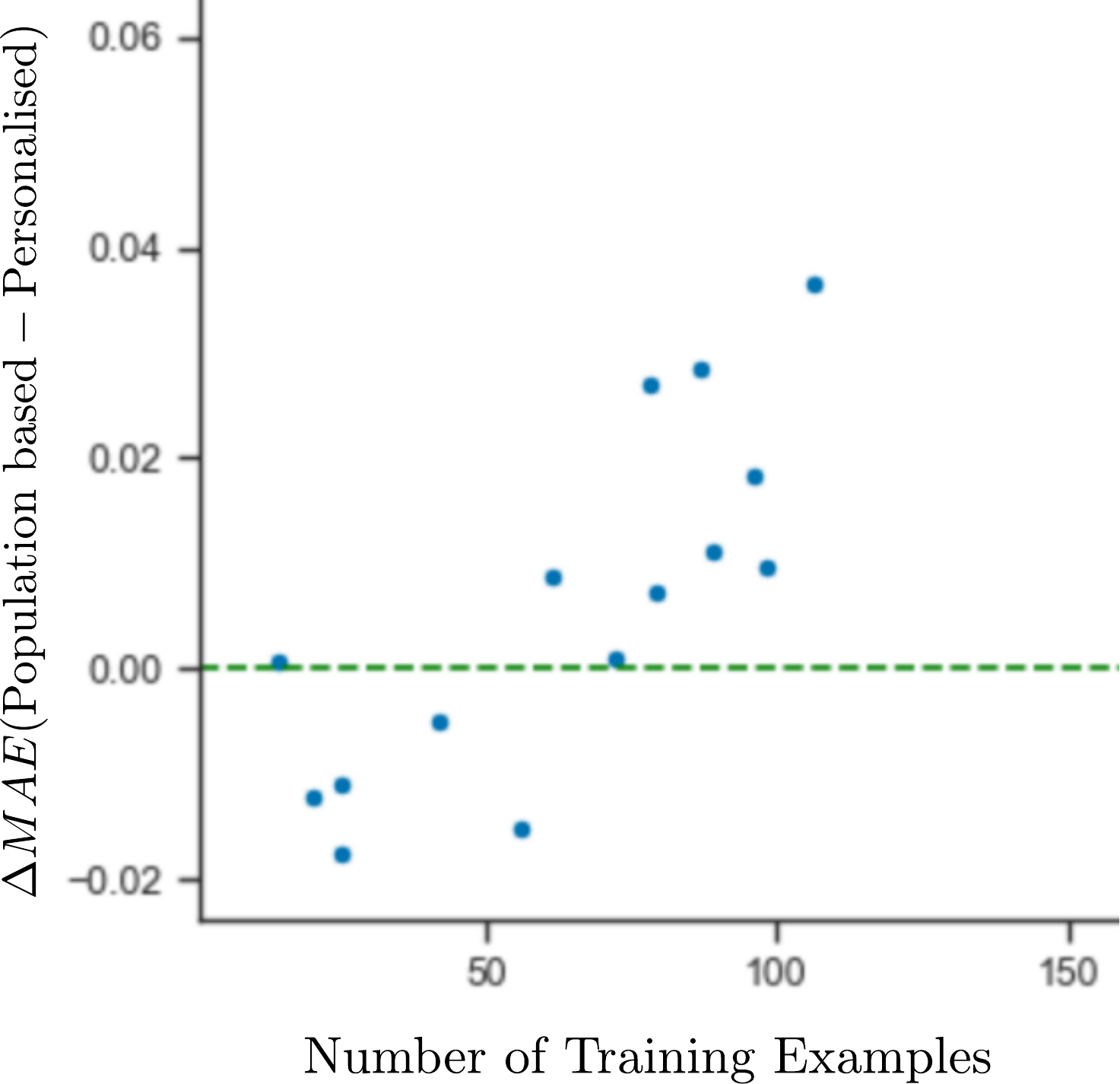}
  \caption{How the difference between Mean Absolute Error (MAE) of population-based and personalised models change with the number of training events per learner. Until about 80 events in this plot, population-based predictions are more accurate. Plot from \cite{context_agnostic_engagement}}\label{fig:person}
\end{figure}

This dataset is uniquely suited for solving the cold-start problem in educational recommenders, both i) \emph{user cold-start}, where new users join the system and we may not have enough information about their context and ii) \emph{item cold-start}, where new educational content is released, for which we may not have user engagement data yet and thus an engagement predictive model would be necessary. The effectiveness of using context-agnostic engagement prediction to address cold-start problem has been empirically demonstrated before (see Figure {\ref{fig:person}}). To the best of our knowledge, this is the largest publicly available dataset to tackle such problems in educational recommenders. The aim of the dataset is not to replace personalised recommenders, but rather complement them providing meaningful population baseline/priors to solve the common cold-start problem. 
While constructing the VLE dataset is a major contribution of this paper, a series of experimental results is also included as an additional contribution. These results validate VLE dataset's usefulness in i) significantly improving engagement prediction models, ii) determining how the training data size impacts model improvements, iv) using the VLN dataset for AI/CS and E-Learning/MOOC educational recommenders and finally, v) showing the feasibility of fusing context-agnostic prediction with personalised recommenders to improve overall prediction performance.

\section{Related Work}

The majority of work in Intelligent Tutoring Systems (ITS) and Educational Recommendation Systems (EduRecSys) revolve around using learner context \cite{kim2021student,truelearn} to predict learner engagement. Although the connection between learner engagement and learning gains have been explored by many \cite{bonafini2017much,Davis18,lan2017behavior}, public datasets in this realm are hard to come by. Many MOOC platforms such as edX \cite{Guo_vid_prod} and Khan Academy \cite{khan_bigdata} harvest valuable data created in an \emph{in-the-wild} setting, yet this data is gated within course owners and consortia\cite{DavisSHH18} (or heavily anonymised) due to its proprietary nature. However, with the boom of online education, understanding features of contents leading to context-agnostic (population-based) engagement, an under researched knowledge area, has become a critical area to explore. This work marks a significant step in this direction by publishing a large dataset that the community can use to push the research frontiers. Other public datasets relating to educational question generation \cite{Chen18,rajpurkar2016squad}, argument strength \cite{persing2015modeling} and essay scoring \cite{AES_dataset} serve different objectives and tasks. 

Study of context-agnostic engagement of video lectures so far has been mainly qualitative, deriving guidelines such as keeping videos short and in parts \cite{Guo_vid_prod} and using conversational language \cite{brame2016effective}. These guidelines are only useful at the content creation stage and have no use in moderating the mammoth of materials already circulating in the Internet. 
Our work, proposes to model context-agnostic engagement using features associated with the educational resource itself, which is useful for scalable quality assurance and recommendation of existing (and newly created) materials.

\subsection{Related Datasets}

A few works on engagement prediction with videos (e.g. modelling watch time) have been done with YouTube \cite{Covington2016,Ribeiro_West_2021}, where YouTube specific meta-data features (e.g. channel reputation, category etc.) are used exclusively. Although large-scale public datasets relating to engagement prediction are encountered, they focus on general-purpose videos (largely entertainment) rather than educational videos \cite{beyondviews}. Some of the features used by these works share similarity with ones proposed in this paper (such as video duration, language and topic features). However, no textual features (based on video transcript) relating to understandability and presentation are used in this prior work, making the methods hard to generalise outside of YouTube. Beyond videos, educational IR \cite{SyedC17,Collins-Thompson:2011} and Wikipedia page quality prediction \cite{Dalip_wiki_svr,wiki_wang} has been attempted using features such as text style, readability, structure, network, recency and review data. Publicly available Wikipedia article quality dataset \cite{Dalip_wiki_svr} with human annotated (explicit) labels is used to tackle the latter task although implicit labels are not included in this dataset. Similar datasets are available for automated essay scoring \cite{taghipour2016neural}. But, none of these datasets fill the lack of resources for predicting engagement of educational videos.

In the context of education, a different line of work looks at learner engagement using learner-specific multi-modal data (brain waives \cite{multa_dataset}, facial expressions \cite{kaur2018prediction} etc.). While tackling a different task, these datasets are mainly collected in a controlled lab setting where the in-the-wild factor is missing \cite{dewan2019engagement}. A large number of public datasets and competitions in education also relate to students interacting with learning problems (e.g. ASSISTments \cite{mendicino2009comparison} or multiple choice questions \cite{choi2020ednet}), but these datasets, contrary to the proposed VLE dataset, do not focus on implicit engagement. More relevant and similar datasets that address context-agnostic engagement prediction in education has been emerging with a focus on MOOCs. Studying approximately 800 videos from edX platform, Guo et al. \cite{Guo_vid_prod} manually processed and provided a qualitative analysis of engagement, with some features being relatively subjective and difficult to automate. A similar work \cite{edx_quality} takes 22 edX videos, extracts cross-modal features and manually annotates their quality with no focus on learner engagement with the videos. Neither dataset is publicly available. MOOCCube is a recently released dataset that contains a spectrum of details relating to MOOC interactions \cite{yu2020mooccube}. Although large, the video watch logs in MOOCCube come from 190,000 users where as VLE signals are generated over 1.1 Million users. As central values (e.g. median) are used for context-agnostic engagement prediction, larger user base adds stability to the engagement centres. MOOCCube uses a closed topic taxonomy disconnected from Wikipedia which prevents the dataset from using all the powerful signals in Wikipedia (e.g. semantic relatedness, category tree to name a few) to improve prediction models. No engagement prediction attempts are published thus far to demonstrate MOOCCube's promise in the task.

On a more relevant contribution, \cite{context_agnostic_engagement} has demonstrated that context-agnostic engagement prediction models can be built with implicit watch time based labels and these models can be used in addressing the cold-start problem (see Figure {\ref{fig:person}}). We identify this work as the closest contribution to the proposed dataset. They gather a collection of 4,000 video lectures with engagement signals generated by 150,000 informal learners. We consider our work, VLE, as an expansion to this dataset with around 12,000 video lectures where engagement signals are generated by 7 times as many learners. The new dataset also restricts itself to strictly English lectures to preserve the meaningfulness of majority of features that are specific to English language. VLE also introduces a new set of Wikipedia-based topical features. Furthermore, this work also accompanies additional experiments that goes beyond data quantity and validates the usefulness of the dataset.

\section{VLE Dataset}

\begin{figure*}[ht]
\begin{center}
    \centerline{\includegraphics[width=1\linewidth]{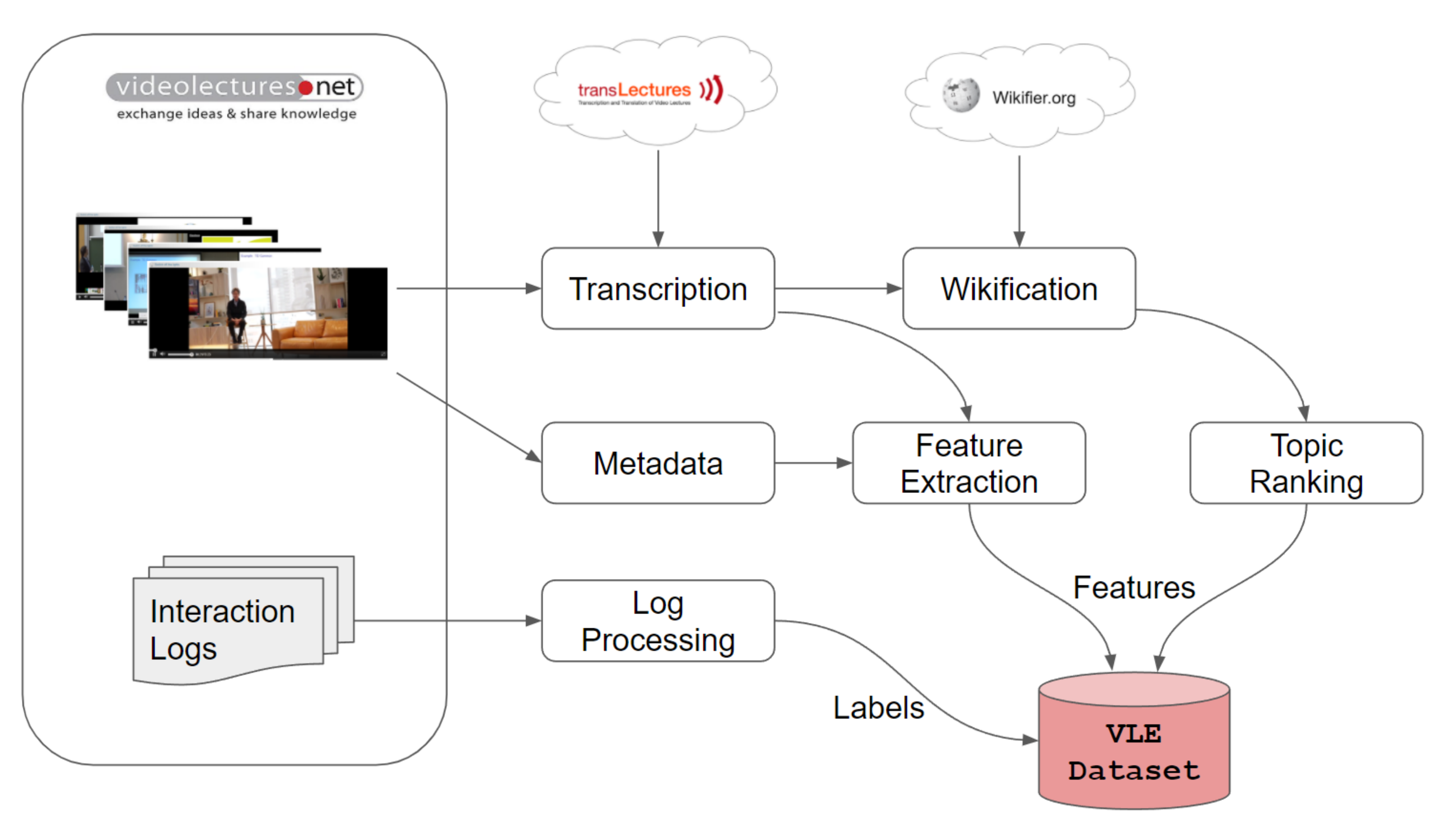}}
    \caption{The video data and the learner interaction logs from VideoLectures.Net repository are processed separately to create the content-based, video-specific features and Wikipedia-based Topics. Multiple different engagement labels are extracted from interaction logs and published in VLE dataset.}
    \label{fig:vle_pipe}
\end{center}
\end{figure*}

VLE dataset is constructed using the aggregated video lectures consumption data coming from a popular OER repository, VideoLectures.Net\footnote{{\url{www.videolectures.net}}} (VLN). These videos are recorded when researchers are presenting their work at peer-reviewed conferences. Lectures are thus
reviewed and material is controlled for correctness of knowledge.  
The presenters and authors of
published 
work that is recorded by VLN agree and provide rights to publish their presentation video, slides and supplementary materials under an open licence in the VLN website which can be used for educational and research purposes.
The majority of research venues covered by VLN is related to Artificial Intelligence and Computer Science, making most videos associating to these topics. Although the dataset mainly consists of scientific talks that are geared towards postgraduate and PhD level learners,
a significant number of tutorials (Table \ref{table:lect_type}) 
are geared for university level students.
In that aspect, many videos in the dataset draw similarities to the style of conventional MOOC lectures. 

The dataset provides a set of statistics aimed at studying population based engagement in scientific videos, together with other conventional metrics in subjective assessment such as 
star ratings and number of views. We believe the dataset will serve the community applying AI in Education to further understand what are the features of educational material that makes it engaging for learners. The users also agree for the user-generated content in the VLN website to be available for research purposes. We have anonymised all user interaction data and aggregated them to ensure privacy of the users. Figure \ref{fig:vle_pipe} gives a high level representation of how different data silos within VLN repository has been used to create the VLE dataset.

\subsection{Feature Extraction} \label{sec:features}

We process the video meta-data and English transcriptions\footnote{provided by \url{www.translectures.eu}.} to extract i) content-based textual features, ii) Wikipedia topic-based features and iii) video-specific features. Majority of the extracted features (with the exception of a few features in the video-specific category) are cross modal (e.g. books, websites and audios) and are easily automatable.

\subsubsection{Content-based Features}
Prior work \cite{context_agnostic_engagement} has proposed a set of effective content-based features for similar datasets. We use the video meta-data and transcript to extract \emph{Word Count} \cite{wiki_wang}, \emph{Title Word Count} and \emph{Document Entropy} \cite{Bendersky2011}, language style features \cite{dalip_quality_features}, \emph{Preposition Rate}, \emph{Auxiliary Rate}, \emph{To Be Rate}, \emph{Conjunction Rate}, \emph{Normalisation Rate}, \emph{Pronoun Rate}, readability related  \emph{Easiness (FK Easiness)}\cite{dalip_quality_features} and language style related \emph{Stop-word Presence Rate}, \emph{Stop-word Coverage Rate} \cite{Bendersky2011,Ntoulas2006}. To represent \emph{Freshness} of lectures (recency), we calculate the number of days between January 01, 1960 and the lecture published date  which is a proxy for recency of the lecture \cite{context_agnostic_engagement}.

\subsubsection{Wikipedia-based Features}

We use Wikifier\footnote{\url{http://wikifier.org}} \cite{wikifier}, a novel entity linking method, on transcripts to extract topical features. Two main types of novel topical features that cover topic authority and topic coverage verticals \cite{quality_features} are proposed through this work.

The \emph{top-5 authoritative topic URLs} and \emph{top-5 PageRank scores} features represent the Topic Authority feature vertical.
Wikifier \cite{wikifier} produces a PageRank score \cite{pagerank} that indicates the marginal authoritativeness of a Wikipedia concept among all Wikipedia concepts associated with a lecture. We use this score to rank and identify the most authoritative topics and use the actual PageRank score as a proxy for quantifying authority. It is noteworthy that \emph{authority} of a learning resource entails author, organisation and content authority \cite{quality_features}. These features represent content authority.

The \emph{top-5 covered topic URLs} and \emph{top-5 cosine similarity scores} features represent \emph{Topic Coverage} feature vertical. The cosine similarity score $cos(s_{tr}, c)$ between the \emph{Term Frequency-Inverse Document Frequency (TF-IDF)} representations of the lecture transcript $s_{tr}$ and the Wikipedia page of concept $c$ is also an output from the Wikifier. We use these values to i) rank and identify most covered topics and ii) quantify the topic coverage \cite{truelearn}. 

\begin{figure}[!tbp] 
  \centering
  \includegraphics[width=\linewidth]{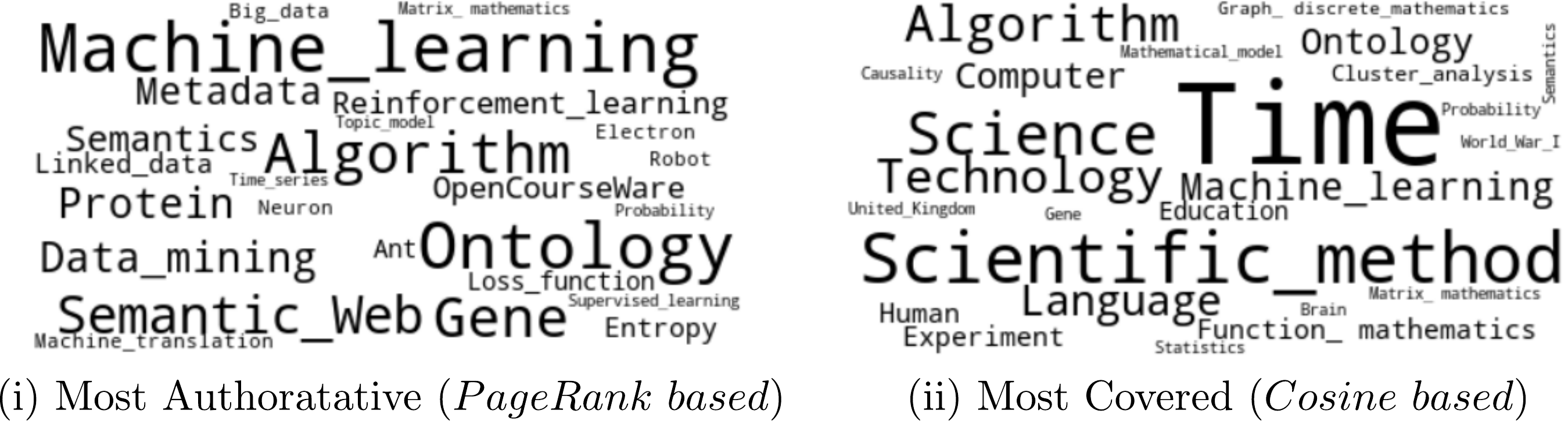}
  \caption{WordClouds summarising the 
  distribution of top 25
  (i) most authoritative and (ii) most covered Wikipedia topics in the dataset. Note that Data Science and Computer Science related topics are most dominant topics.}
  \label{fig:wordcloud}
\end{figure}

For authority and coverage, the top 5 topic URLs and their scores are included introducing four additional feature groups providing 20 distinct feature columns in the final dataset. Figure {\ref{fig:wordcloud}} presents two word clouds that show the 25 most authoritative and covered topics in the VLE dataset. With respect to the topics in a lecture, the authoritative topics are the ones that have strong connection (linkage in Wikipedia) with other co-occurring topics in a lecture
whereas the covered topics are the ones with high textual overlap with Wikipedia concept pages. This figure confirms that these two feature groups represent different things although there is some overlap between the topics.

\subsubsection{Video-specific Features} 
We identify a set of easily automatable, prior proposed \cite{context_agnostic_engagement} video specific features. \emph{Lecture Duration, Is Chunked, Lecture Type}\cite{Guo_vid_prod}, \emph{Silence Period Rate (SPR)} and \emph{Speaker Speed} \cite{context_agnostic_engagement} are calculated based on prior work . \emph{Lecture Duration} is reported in seconds. \emph{Is Chunked} is a binary feature which indicates if a lecutre has multiple parts. \emph{Lecture type} value is derived from the metadata. The possible values for this feature are described in Table \ref{table:lect_type}. 

Table \ref{table:lect_type} also gives insight into how diverse the VLE dataset is. There are  different types of videos such as research presentations (e.g. $\texttt{vl,vps,vpr,}\dots$ ), scientific talks (e.g. $\texttt{vbp,vdm,vkn,vit,}\dots$), dialogues (e.g. $\texttt{vpa,vdb,}\dots$) and tutorials (e.g. $\texttt{vtt}$) among the dataset. 

\begin{table}
\small \centering \setlength{\tabcolsep}{4pt}
\caption{14 types of lectures in the VLE dataset and their abbriviation (Abbr.) and frequency (Freq).}
\label{table:lect_type}

\begin{tabular}{ l l r l l r} 
    \hline
    Abbr.&Description&Freq.&Abbr.&Description&Freq.\\ 
    \hline
    \texttt{vbp}&Best Paper&67&\texttt{vdb}&Debate&135 \\
    \texttt{vdm}&Demonstration&315&\texttt{viv}&Interview&121 \\
    \texttt{vid}&Introduction&75&\texttt{vit}&Invited Talk&609 \\
    \texttt{vkn}&Keynote&274&\texttt{vl}&Lecture&7125 \\ 
    \texttt{vop}&Opening&190&\texttt{oth}&Other&58 \\
    \texttt{vpa}&Panel&207&\texttt{vps}&Poster&162 \\ 
    \texttt{vpr}&Promotional Video&69&\texttt{vtt}&Tutorial&2142 \\  
    \hline
\end{tabular}
\end{table}

\subsection{Labels}

Multiple labels based on explicit and implicit feedback are provided with this dataset which will allow researchers to compare between different labels and also integrate multiple label types when modelling engagement. Three main types of quantification of engagement labels are presented.

\subsubsection{Explicit Ratings} \emph{Mean Star Rating} based on a rating scale of 1-5 for each lecture is provided. This value is accompanied by the number of ratings used to calculate the mean. The proposed dataset has 2,127 ratings (almost 2x than Bulathwela et al. \cite{context_agnostic_engagement}). Missing ratings are labelled with \texttt{-1}.

\subsubsection{Popularity} The total number of views, named \emph{View Count}, for each video lecture as of February 1, 2021 is extracted from the metadata and provided with the dataset.

\subsubsection{Watch Time/Engagement} The majority of learner engagement labels in the VLE dataset are based on {watch time}. We compute Normalised Engagement Time (NET) to compute the \texttt{Median of Normalised Engagement (MNET)}, as it has been proposed as the gold standard for engagement with educational materials in previous work \cite{Guo_vid_prod}. We further compute \texttt{Average of Normalised Engagement (ANET)}.
To have the MNET and ANET labels in the range $[0,1]$, we set the upper bound to 1 and derive Saturated MNET (\emph{SMNET}) and Saturated ANET (\emph{SANET}) that are included in the dataset.

The standard deviation of \texttt{NET} for each lecture (\emph{Std of Engagement}) is reported, together with the \emph{Number of User Sessions} used for calculating \texttt{MNET} and \texttt{ANET}. These measure allows understanding stability of the centres published. The set of individual NET values are also published to allow future researchers to exploit the true distribution of values.

\subsection{Preserving Anonymity and Ethics} \label{sec:ethics}
We only publish lectures with more than 5 views to preserve k-anonymity and avoid revealing learner identities \cite{orcas_dataset}. A regime of additional techniques are used to preserve lecturer anonymity in order to avoid having unanticipated effects on lecturer's reputation by associating implicit learner engagement values to their content.

Rarely occurring \emph{Lecture Type} values were grouped together to create the \emph{other} category in Table \ref{table:lect_type}. 
Similarly, subject categories Life Sciences, Physics, Technology, Mathematics, Computer Science, Data Science and Computers subjects to \texttt{stem} category and the other subjects to \texttt{misc} category. Rounding is used with \emph{Freshness} and \emph{Lecture Duration} to the nearest 10 days and 10 seconds respectively. Gaussian white noise (10\%) is added to \emph{Title Word Count} feature and rounded to the nearest integer.
 
VLE dataset comes from a video lecture collection that mainly belongs to the Computer Science community, where there is a gender imbalance, both in audience and presenters. To enhance the neutrality of our findings and contributions, we have avoided using feature classes that could potentially reflect gender characteristics. For example, we have avoided using visual features (facial features, presenter emotions) and sound related features (pitch) that may actively or passively embed gender characteristics. Instead we have focused on features that mainly reflect informational content of the lectures. Furthermore, where video specific features are incorporated, we have used very generic features such as "speaker speed" that are unlikely to be correlated with characteristics such as gender or age.
 \subsection{Final Dataset}
 The final dataset includes lectures that are 
 published between September 1, 1999 and December 31, 2020. The 
 engagement labels are created from events of over 1.1 Million users logged between December 01, 2016 and February 01, 2020. 
 The final dataset contains 11,548 lectures across 21 subjects (eg.~Computer Science, Philosophy, etc. with a majority from AI and Computer Science) that are grouped into STEM and Miscellaneous categories. The collection of videos span various video lengths with the duration distribution having two modes at approx. 2000s (33 mins) and 4000s (1hr) time points which align with typical lengths of research talks and presentations. The mean word count of the videos is 5347.9. The video lecture collection uses on average 93.9 learners per video when calculating engagement centres. The dataset, helper tools, example code snippets and various descriptive statistics related to the VLE dataset are available publicly\footnote{\url{https://github.com/sahanbull/VLE-Dataset}}. 

\subsection{Supported Tasks} \label{sec:taks}

This section introduces the reader to the tasks that the dataset could be used for. The main application areas of these tasks are i) quality assurance in educational video repositories and understanding and ii) predicting context-agnostic engagement in an web-based learning setting. 
We establish two main tasks, which we mainly focus on in this paper, that can be objectively addressed using the VLE dataset using a supervised learning approach. These are:
\begin{itemize}
    \item \textbf{Task 1: Predicting context-agnostic (population-based) engagement of video lectures}: The dataset provides a set of relevant features and labels to construct machine learning models to predict context-agnostic engagement in video lectures. 
    The task can be treated as a regression problem.
    \item \textbf{Task 2: Ranking of video lectures based on engagement}: Building predictive models that can rank videos based on their context-agnostic engagement could be useful in the setting of an educational recommendation system, including tackling the cold-start problem associated to new video lectures. The task can be treated as a ranking problem to predict the global/relative ranking of video lectures. 
\end{itemize}

\paragraph{Other Tasks}  Several auxiliary tasks  can also be addressed with this dataset. This dataset is suitable for, not limiting to, several tasks such as i) understanding influential features for engagement prediction ii) understanding the strengths and weaknesses of different implicit/explicit labels, that have been investigated in prior work with similar datasets \cite{context_agnostic_engagement,perez2019pairwise}. 
The video representations, with the use of unsupervised approaches can be used to understand meaningful hidden patterns contrasting between clusters of videos (e.g. talks vs. lectures vs. tutorials). 
With the use of Wikipedia based topics, we also see opportunities in deducing the structure of knowledge based on how topics co-occur within videos. Such investigations can be done on this dataset in isolation or can be fused with similar datasets where this task has been attempted before \cite{yu2020mooccube}.  

\subsection{Evaluating Performance}
 We identify \emph{Root Mean Squared Error (RMSE)} as a suitable metric for evaluating Task 1. Measuring RMSE against the original labels published with the datasets will allow different works to be compared fairly. With reference to Task 2, we identify \emph{Spearman's Rank Order Correlation Coefficient (SROCC)} as a suitable metric. 
 SROCC 
 is suitable for comparing between ranking models that create global rankings
 (e.g. point-wise rankers). 

 We use 5-fold cross validation to evaluate model performance with tasks 1 and 2. The folds are released together with the dataset, to allow to facilitate fair comparison and reproducability. The five folds can be identified using the \texttt{fold} column in the dataset. 
5-fold cross validation allows reporting the \emph{confidence intervals (1.96 $\times$ Standard Error)} of the performance estimate, which we include in Table \ref{tab:accuracy}.
 
\section{Baselines and Experiments} \label{sec:baselines}
Through our experiments
we seek answers to multiple research questions, which we detail below. Note that these research questions overlap only partially with the proposed supported tasks outlined in section \ref{sec:taks}, as the use purposes of the dataset go much beyond what we could explore in this paper. The main research questions of interest are: 
\begin{itemize}
    \item \textbf{RQ1: } Does the newly constructed VLE dataset lead to training more performant prediction models?
    \item \textbf{RQ2: } How does the larger quantity of training data affect predictive performance?
    \item \textbf{RQ3: } Is the model useful for modelling engagement with Computer Science materials?
    \item \textbf{RQ4: } Is this dataset useful for modelling engagement in E-Learning lectures and MOOC videos?
    \item \textbf{RQ5: } Does context-agnostic engagement prediction help in the cold-start scenario?
\end{itemize}

Prior work by Bulathwela et al. \cite{context_agnostic_engagement} demonstrated \emph{Random Forest (RF)} model obtains best performance among linear and non-linear models in similar datasets. Therefore, we use the RF model to benchmark the new VLE dataset for Tasks 1 and 2 described earlier. In addition, a handful of \emph{Multi Layer Perceptron (MLP)} architectures were also experimented with due to their success in engagement prediction with YouTube videos \cite{Covington2016}. 

The code for extracting the proposed features is also published with this work in the dataset repository.

\subsection{Labels and Features for Baseline Models}

\emph{SMNET} label is used as the target variable for both tasks. Preliminary investigations indicated that SMNET label follows a Log-Normal distribution, motivating us to use a log transformation on the SMNET values before training the models. Empirical results further confirmed that this step improves the final performance of the models. We undo this transformation for computing $RMSE$ while this transformation doesn't affect $SROCC$.

All the features outlined as the content-based and video-based sections in section \ref{sec:features} are included in the baseline models. 
The models are trained with three different feature sets in an incremental fashion: 
\begin{enumerate}
    \item \emph{Content-based}: Features extracted from lecture metadata and the transcript-based textual features. 
    \item \emph{+ Wiki-based}: Content-based + 2 Wikipedia-based features (Top 1 Most Authoritative Topic URL and Most Covered Topic URL).
    \item \emph{+ Video-based}: Content-based + Wikipedia-based + Video-specific features.
\end{enumerate}

However, due to the large amount of topics in the Wikipedia-based feature groups, we restrict to the top 1 authoritative and covered topic features where they are encoded as binary categorical variables. Our initial attempts to encode these features in a reduced dimension space (using Singular Value Decomposition) led to deteriorated results contrary to our expectations. Practitioners are encouraged to try further encoding of the topic variables, as it will likely have a positive impact on the performance.

\subsection{Experiments}

Addressing RQ1, both the RF and MLP models are experimented with the proposed features sets with the smaller $\texttt{4k}$ dataset \cite{context_agnostic_engagement} and the newly proposed VLE dataset. 5-fold cross validation is used in this experiment. This setup allows identifying i) how performance gains are achieved through adding each new group of features and ii) how performance gains are achieved through adding new observations. Follow on experiments addressing RQ2 and 3 are run using folds 1-4 as training data and fold 5 of the dataset as testing data. We experiment by using varying proportions of training data to train the model. When selecting training data, random sampling is used to keep the diversity of the lectures similar to the full dataset. All the trained models using different quantities of training data are then evaluated on the same held out test set. 

To validate RQ4, we first partition the entire dataset to two parts, i) tutorial videos (\texttt{vtt} in Table \ref{table:lect_type}) and ii) all other videos, as test and train data respectively. However, tutorials presented in a research conference may significantly vary from e-learning videos geared for course learning. To address this mismatch, we further identify 1,035 videos (among the tutorials) that exclusively belong to the Open Course Ware Consortium (OCWC)\footnote{\url{http://videolectures.net/ocwc}}. OCWC contains university lectures that have been intended for course teaching via e-learning. These lectures are recorded using a variety of MOOC production techniques such as classroom lecture, talking head and power point presentation \cite{Guo_vid_prod}. We define these videos as \texttt{ocw} lectures. We use the training data (all except tutorials) to train the engagement model and evaluate prediction performance on i) OpenCourseWare, \texttt{ocw} lectures, ii) all tutorials but OpenCourseWare, \texttt{!ocw} and iii) all tutorials \texttt{vtt}, (Entire test data). The follow on experiments (RQ2-4) are only done with the best performing model from Table \ref{tab:accuracy} (RF model with \emph{Content + Wiki + Video} feature group) to reduce computational cost. 

A different experiment was run to answer RQ5. We utilise TrueLearn Novel \cite{truelearn} (hereby referred to as \emph{TrueLearn}), a recently proposed educational recommender model that learns individualised models to predict engagement with video lectures. A key limitation with personalised models such as TrueLearn is that there is no information about the user in the beginning, leading to a user cold-start problem which effectively means having ill-informed engagement predictions in the beginning of the learner session. To address this, we utilise the proposed context-agnostic engagement prediction model where a hybrid recommendation system (hereby referred to as \emph{TrueLearn$_{++}$}) is built by combining it with TrueLearn model. For simplicity's sake, TrueLearn$_{++}$ uses "switching" \cite{burke2002hybrid}, where the context agnostic model is used to make a prediction for the \emph{first} event of the user (where the personalistion model has no information to work with) and then switched to TrueLearn model which can exploit the user watch history. PEEK dataset \cite{peek_orsum}, which includes more than 20,000 user sessions, is used for the experiment where the context agnostic model is trained using lectures that are not present in the PEEK test data. Then the predictive performance on the PEEK test data using TrueLearn (The baseline) and TrueLearn$_{++}$ (where the first event is predicted using the context-agnostic model) is measured using Accuracy, Precision, Recall and F1-Score. To evaluate if the improvement of metrics is statistically significant, a learner-wise, one-tailed paired t-test is used.
\section{Results and Discussion}

\begin{table}[!tbp]
\centering
\scriptsize \setlength{\tabcolsep}{4pt}
\caption{RMSE and SROCC with confidence intervals for the engagement prediction (Task 1) and lecture ranking (Task 2) using the Random Forests model with both \texttt{4k} \cite{context_agnostic_engagement} and \texttt{12k} (Our VLE) datasets. Better performance per task is highlighted in \textbf{bold}.}\label{tab:accuracy}
\begin{tabular}{l | l l | l l } \hline
& \multicolumn{2}{c}{$RMSE$ with {Task 1}}  & \multicolumn{2}{c}{$SROCC$ with {Task 2}}  \\
Feature set  & \multicolumn{1}{c}{\texttt{4k}} & \multicolumn{1}{c}{\texttt{12k} \emph{(Ours)}} & \multicolumn{1}{c}{\texttt{4k}} & \multicolumn{1}{c}{\texttt{12k} \emph{(Ours)}} \\
\hline
Content-based & {.1801$\pm$.006} & \textbf{.1170$\pm$.006} & 
.6190$\pm$.011 & \textbf{.7504$\pm$.013} \\
+ Wiki-based & {.1798$\pm$.007} & \textbf{.1178$\pm$.006} &
{.6251$\pm$.014} & \textbf{.7505$\pm$.013} \\
+ Video-specific & {.1728$\pm$.007} & \textbf{.1098$\pm$.007} & 
.6758$\pm$.020 & \textbf{.7832$\pm$.009}\\
\hline

\end{tabular}
\end{table}

\begin{figure}[!tbp] 
  \centering
  \includegraphics[width=1.025\linewidth]{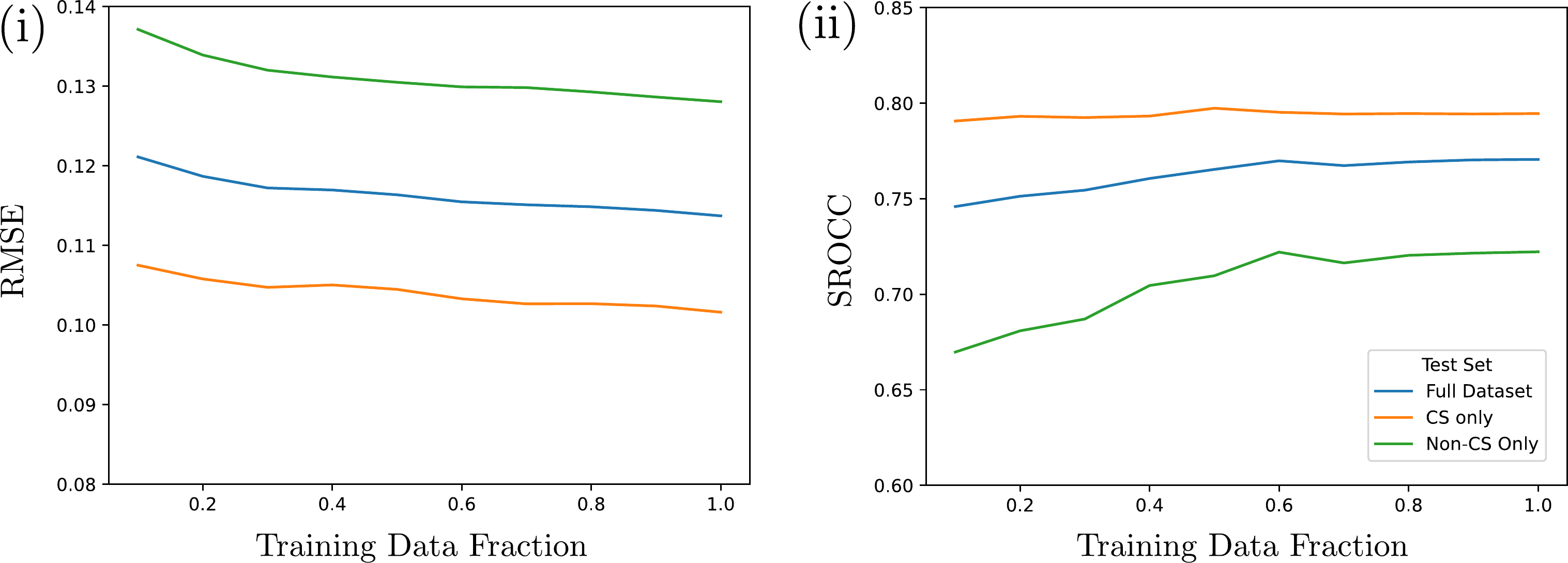}
  \caption{Predictive performance for (i) engagement prediction and (ii) lecture ranking tasks with varying proportions of randomly sampled training data. The test set performance for full test dataset (Blue) and subsets of test dataset that consists of CS lectures only (Orange) and Non-CS lectures only (Green) are also reported}\label{tab:ocw}
  \label{fig:train_size}
\end{figure}

\begin{table}[!tbp]
\centering \small \setlength{\tabcolsep}{5pt}
\caption{Performance for OpenCourseWare (\texttt{ocw}), Non-OpenCourseWare  (\texttt{!ocw}) tutorial and All tutorial (\texttt{vtt}) videos for engagement prediction and lecture ranking tasks. Better performance per task is highlighted in \textbf{bold}.}
\begin{tabular}{l c c c c}
\hline
 & \texttt{ocw} & \texttt{!ocw} & \texttt{vtt} & From Table \ref{tab:accuracy}\\
\hline
$RMSE$ with Task 1 & .0539 & \textbf{.0404} & .0406 & .1098\\
$SROCC$ with Task 2 & \textbf{.9485} & .9209 & .9223 & .7832\\
\hline
\end{tabular}
\end{table}

\begin{table}[!tbp] 
\centering 
\caption{Average test set performance for Accuracy (Acc.), Precision (Prec.), Recall (Rec.) and F1-Score (F1). The more performant value is highlighted in \textbf{bold}. The metrics where the proposed model that outperform the baseline counterpart in the \texttt{PEEK} dataset ($p< 0.01$ in a one-tailed paired t-test) are marked with $\cdot^{(*)}$.}
\label{tab:personalise}
\begin{tabular}{lllll}
\hline
\multicolumn{1}{c}{Model} & Acc. & Prec. & Rec. & F1 \\
\hline
Truelearn & 62.69 & 57.54 & \textbf{81.88} & \textbf{64.98} \\
Truelearn$_{++}$ & \textbf{63.51}$\cdot^{(*)}$ & \textbf{57.91}$\cdot^{(*)}$ & 79.13 & 64.39 \\
\hline
\end{tabular}
\end{table}

\begin{table}[!tbp]
\centering 
\caption{Test set performance for Accuracy (Acc.), Precision (Prec.), Recall (Rec.) and F1-Score (F1) for \emph{first} event of each learner. The more performant value is highlighted in \textbf{bold}. The metrics where the proposed model that outperform the baseline counterpart in the \texttt{PEEK} dataset ($p< 0.01$ in a one-tailed paired t-test) are marked with $\cdot^{(*)}$.}
\label{tab:event1}
\begin{tabular}{lllll}
\hline
Model & Acc. & Prec. & Rec. & F1 \\
\hline
TrueLearn & 44.21 & 44.21 & \textbf{100.00} & \textbf{61.32} \\
TrueLearn$_{++}$ & \textbf{56.09}$\cdot^{(*)}$ & \textbf{50.32}$\cdot^{(*)}$ & \ \ 53.58 & 51.90 \\
\hline
\end{tabular}
\end{table}

The performance metrics observed with the RF model on Task 1 and 2 (RQ1) are outlined in Table \ref{tab:accuracy}. Although we expected competitive results with MLP models, we failed to observe promising results. We believe this is due to the restrictive architecture space we used in our experiments. We encourage future researchers to experiment with more complex and wider range of neural architectures that may lead to better results. 

Figure {\ref{fig:train_size}} illustrates how the training data size impacts the i) RMSE and ii) SROCC (RQ2). It also demonstrate the performance difference between predicting engagement of Computer Science videos (CS) vs. Non-CS videos (RQ3). Finally, Table \ref{tab:ocw} presents predictive performance of the model on e-learning type lectures and tutorials (RQ4).

The overall performance results relating to the effect of \emph{combining} the context agnostic model with personalisation models to battle cold-start problem (RQ5) is reported in Table \ref{tab:personalise}. A magnified view of the performance of predicting the outcome of the first event of each user (where the context-agnostic predictor is supposed to help the personaliser) is reported in Table \ref{tab:event1}.

\subsection{Performance Gains and Causes (RQ1-2)} 

Results in Table \ref{tab:accuracy} clearly shows that the 300\% larger VLE lead to significant performance gains in both engagement prediction and lecture ranking tasks. In the case where the full feature set is used with the RF model, RMSE on Task 1 drops from  .17 to .1 (41\%) while SROCC on Task 2 jumps from .68 to .78 (15\%). The labels (both explicit and implicit) themselves are more accurate as they are calculated using a larger user population. This means that many of the lectures that already existed in the smaller dataset \cite{context_agnostic_engagement} are likely to get improved engagement labels as the new labels are calculated using more user sessions that interacted with the videos during a wider time period (including very recent sessions until February 2020). 

Within the VLE dataset itself, using additional feature groups tend to lead to better performing models. Results for VLE dataset in Table \ref{tab:accuracy} demonstrates this trend where a significant jump in performance is evidenced when incorporating modality specific features (Video-specific features). However, the results also show that the cross-modal content-based features alone lead to substantial performance. This is a good indication that easy-to-compute, cross-modal features alone are sufficient to build a system that can predict context-agnostic engagement of videos. In a practical viewpoint, the proposed cross-modal features are computationally light (unlike complex deep models, e.g. vision models). Wikification, used in generating Wiki-based features, also operates at web-scale\footnote{\url{http://wikifier.org}}.
Although the results show minute gains by adding the Wiki-based features, we believe that this is due to the simplicity of the Wiki features used in constructing the baselines leaving much room for sophistication (e.g. exploiting the semantic relatedness of topics). The topics, coming from a humanly-intuitive taxonomy, leaves room for building interpretable features. 

Figure {\ref{fig:train_size}} confirms that the increase of training data is leading to performance gains in both tasks. Figure {\ref{fig:train_size}(i)} suggests that the Root Mean Square Error (RMSE) can be further improved with more training examples. However, Figure {\ref{fig:train_size}(ii)} tells rather a different story where the performance gain seem to saturate around 60\% mark. This is an indication that improving ranks of the test data gets significantly harder at around  5,500 training examples ($\approx$ 60\% of the 9,239 training set) .

\subsection{Relevance to AI/CS Education (RQ3)}

The follow up experimental results in Figure {\ref{fig:train_size}} demonstrates that this dataset allows achieving higher performance on CS-only lectures. A likely reason for this may be the higher diversity of lecture in the non-CS category as it consists of significantly different subjects. Nevertheless, Figure {\ref{fig:train_size}} shows that a test set RMSE of $\approx .1$ and SROCC of $\approx .8$ is achievable with CS lectures. Figure {\ref{fig:wordcloud}} further shows that majority of the lectures in the dataset contains concepts relating to Artificial Intelligence and Data Science (e.g. Machine Learning, Ontology, Semantic Web ...). This indicates that majority of the CS lectures in the dataset contain topics relating to  AI making this dataset a highly suitable dataset for training engagement prediction models for AI and CS education.

\subsection{Relevance to E-Learning and MOOCs (RQ4)}

Table \ref{tab:ocw} shows strong evidence that the models trained with VLE dataset generalise really well for engagement modelling in e-learning type videos created for course teaching amid the dataset containing many different video types (as per Table \ref{table:lect_type}). The models trained are much better at engagement prediction and ranking of tutorial-like videos than general scientific talks. Having tested with lectures that have been recorded using different MOOC video production techniques, the high performance obtained on \texttt{ocw} lectures confirms that VLE dataset can be highly effective in building context-agnostic engagement models for e-learning and MOOC systems.

\subsection{Relevance to Addressing Cold-Start (RQ5)}

Table \ref{tab:personalise} shows that by simply incorporating the context-agnostic engagement prediction in TrueLearn Novel algorithm (together becoming TrueLearn$_{++}$) can lead to significant improvements in accuracy and precision. The same table also shows that the drop of overall F1 score can be attributed to the steep drop of Recall Score. Table \ref{tab:event1} sheds more light into where this steep drop of recall occurs. This is, the baseline TrueLearn model always predicts positive engagement for the first event of the user. As seen in table \ref{tab:event1}, the recall of baseline TrueLearn model being 1.0 while the accuracy and precision being the same depicts this fact. TrueLearn predicts positive in event 1 of each user because the model has no information to base the prediction on \cite{truelearn}. However, table \ref{tab:event1} shows that the scenario is different in TrueLearn$_{++}$ as the model has additional information during the first event. Both accuracy and precision of predictions in first event of the learner population significantly improves. The recall will fall as the proposed context-agnostic model only captures a population based prior which may deviate from the individuality of the learners. However, it can be argued that making a prediction with additional information is better than predicting with no prior information. In the bigger picture, being able to make slightly more informed and varied predictions for the first event of learners based on lecture content features enable significantly improving prediction accuracy and precision of TrueLearn$_{++}$ in Table \ref{tab:personalise}. It is also noteworthy that our experiment, for the sake of simplicity, uses a rule that could be significantly improved further, e.g. using weights of the probabilities of both population-based and personalised models at the beginning of a user session (using weighting or stacking \cite{burke2002hybrid}), where the weight of population-based engagement decreases as we gather more information about the user.
as we gather more information about the user.

\section{Opportunities and Limitations} \label{limitations}

The VLE dataset brings plenty of opportunities to the research community that is interested in building context-agnostic engagement using content related features. It is significantly larger (15x than \cite{Guo_vid_prod} and 3x than \cite{context_agnostic_engagement}) and more education focused than the contenders \cite{beyondviews} focused on engagement with videos. The larger quantity of examples may even enable more complex model families (e.g. deep learning) to be used in this task although our limited early experiments did not reap fruit. Another noteworthy opportunity is that this dataset can be extended periodically to create even larger and more accurate evaluations of the dataset. 

The Wiki-based features open up limitless possibilities as many sophisticated feature sets can be built and experimented. Due to the connectivity to Wikipedia, both its content and link structures can be exploited to invent meaningful, yet interpretable features. A further step can enable other data structures such as knowledge bases (e.g Wikidata), category tree etc. to be used for feature creation.

Support for other tasks beyond the main tasks will allow this dataset to be useful in creating decision support systems that can help future content creators and help them create engaging educational videos \cite{kurdi2021think} while confirming prior findings \cite{Guo_vid_prod,brame2016effective,context_agnostic_engagement} regarding drivers of learner engagement. Furthermore, the dataset also enables comparing the findings from studies outside of education \cite{beyondviews}. The Wikipedia features also facilitates grounds to explore topic related tasks such as learning the structure of knowledge \cite{yu2020mooccube}.

There also exists limitations in this dataset. VLE dataset is largely comprised of Computer Science and Data Science materials (Figure \ref{fig:wordcloud}) that are delivered all in English. While this is an opportunity for AI and Computer Science education, results in Figure {\ref{fig:train_size}} also shows that this fact leads to comparatively \emph{less fruitful} non-CS results. The dataset and its features are also not suitable for non-English video collections. Amid its size, the dataset still lacks variety of materials in topical and lingual sense.

At first sight, the majority of the lectures in our dataset come from male presenters, potentially creating a significant gender imbalance in the dataset. As pointed out in section \ref{sec:ethics}, we have taken some measures to restrict the feature set to what we believe to be more neutral features. However, since we do not have access to gender information in the data collected, it is impossible to test and guarantee that there is zero correlation between the proposed features in the VLE dataset and negative gender biases. Care should be taken when enhancing these features and there is room to do more rigorous tests to understand if any gender biases are present within the dataset.

\emph{Learner Engagement} is a loaded concept with many facets. In relation to consuming videos, many behavioural actions such as pausing, rewinding and skipping can contribute to latent engagement with a video lecture \cite{lan2017behavior}. 
Due to the technical limitations of the platform and privacy concerns, only watch time, view and mean ratings are included in this dataset. Although watch time has been used as a representative proxy for learner engagement with videos \cite{Guo_vid_prod,beyondviews,Covington2016}, we acknowledge that more informative measures may lead to more complete engagement signals.

\section{Conclusions}

Identifying the need of resources to push the frontiers of context-agnostic engagement prediction, which is a crucial part of addressing scalable quality assurance and the cold-start problem in educational recommenders, we release the VLE dataset. This dataset consists of around 12,000 videos with three groups of features, namely, i) content-based, ii) Wikipedia-based and iii) Video-specific features, accompanied as well by iv) multiple implicit and explicit engagement labels. We establish 2 main tasks, and identify multiple other tasks that can be tackled with this dataset. Addressing the 2 main tasks proposed, we benchmark the new dataset to show significant prediction gains over a similar yet, smaller prior dataset. In follow on experiments, we investigate how the magnitude of training examples relate to performance gains while also demonstrating the suitability of this dataset to build models for AI/CS related video collections. We further validate that the dataset can be used to model engagement with e-learning type video lectures to show its relevance to educational recommendation systems in the context of MOOCs. With the use of a simple experiment, it is also demonstrated that such a model can be incorporated with a personalised (contextual) engagement prediction model to significantly improve the predictions.

\paragraph{Future Directions} 
We see several lines of future work addressing current limitations of the dataset. Adding diverse examples from different subject areas is our top priority. Experimenting further with more sophisticated Wiki-based features (e.g. by exploiting the Wikipedia semantic graph \cite{ponza2020computing}) is a promising way forward. The possibility of including more learner engagement related signals (e.g.: pauses, replays, skips, etc.) will be explored in the subsequent version of the dataset, without compromising learner privacy.As more understanding of engagement with other modalities (such as PDFs and e-Books) is gained, it is possible to add more learning resources from diverse modalities to widen the horizons of the dataset and improve understanding of engagement with different modalities of educational material. 

This work only devices a simple mechanism to combine the context-agnostic model with a personalisation model for the sake of proving that the proposal has genuine use-cases. The experiment barely scratches the surface on how a context-agnostic engagement prediction model can be incorporated in an educational recommendation system. There are a variety of alternatively  and more sophisticated approaches that can potentially have bring larger performance gains, which we will explore in future work.

\section{Acknowledgments}
This research was partially conducted as part of the X5GON project funded from the EU's Horizon 2020 research programme grant No 761758.
This work is also supported by the European Commission funded project "Humane AI: Toward AI Systems That Augment and Empower Humans by Understanding Us, our Society and the World Around Us" (grant 820437) and the EPSRC Fellowship titled "Task Based Information Retrieval" (grant EP/P024289/1). 

%
\bibliographystyle{abbrv}
\bibliography{edm22}  
\end{document}